\documentclass[prl,letterpaper,reprint]{revtex4-1}
\usepackage[utf8x]{inputenc}
\usepackage{amsmath}
\usepackage{graphicx}% Include figure files
\usepackage{dcolumn}% Align table columns on decimal point
\usepackage{bm}% bold math
\usepackage{subfigure}

\newcommand{\ket}[1]{\left| #1 \right\rangle}
\newcommand{\bra}[1]{\left\langle #1 \right|}

\newcommand{\abs}[1]{\left| #1 \right|}
\newcommand{\dpar}[2]{\frac{\partial #1}{\partial #2}}

\newcommand{\dx}[1]{{\rm d} #1 \,}

\begin{document}
\begin{abstract}
The bulk photovoltaic effect is a long-known but poorly understood phenomenon.  Recently, however, the multiferroic bismuth ferrite has been observed to produce strong photovoltaic response to visible light, suggesting that the effect has been underexploited as well. Here we present three polar oxides in the LiNbO$_3$ structure that we predict to have band gaps in the 1-2~eV range and very high bulk photovoltaic response: PbNiO$_3$, Mg$_{1/2}$Zn$_{1/2}$PbO$_3$, and LiBiO$_3$.  All three have band gaps determined by cations with $d^{10}s^0$ electronic configurations, leading to conduction bands composed of cation $s$-orbitals and O $p$-orbitals.  This both dramatically lowers the band gap and increases the bulk photovoltaic response by as much as an order of magnitude over previous materials, demonstrating the potential for high-performing bulk photovoltaics.
\end{abstract}
\title{\bf First-principles materials design of high-performing bulk photovoltaics with the LiNbO$_3$ structure} 
\author{Steve M. Young}
\affiliation{Center for Compuational Materials Science, United States Naval Research Laboratory, Washington, DC 20375, USA}
\author{ Fan Zheng and Andrew M. Rappe}
\affiliation{The Makineni Theoretical Laboratories, Department of Chemistry, University of Pennsylvania, Philadelphia, Pennsylvania 19104-6323, USA}

\maketitle
Photovoltaic effects have long been observed in bulk polar materials, especially ferroelectrics~\cite{Chynoweth56p705,Chen69p3389,Glass74p233,Fridkin01p654}. Known as the bulk photovoltaic effect (BPVE), it appeared to derive from inversion symmetry breaking.  Despite intense initial interest, early explorations revealed low energy conversion efficiency, in part due to the high band gaps of most known ferroelectrics.  Additionally, despite several proposed mechanisms, the physical origin of the BPVE was unclear~\cite{Chen69p3389,vonBaltz81p5590,Sturman1992,Jermann93p2085,Tonooka94p224}.  

However, recent emphasis on alternative energy technologies and the observation of the effect in novel semi-conducting ferroelectrics (with band-gaps in the visible range) has renewed interest~\cite{Grinberg13p509,Vijayaraghavan14p6530,Somma14p146602,Furchi14p4785,Zenkevich14p161409,Zheng15p31}.  Several studies have attempted to elucidate the various contributions to the photovoltaic response -- bulk or otherwise -- in ferroelectrics~\cite{Ichiki05p222903,Qin09p022912,Yue10p015104,Yuan09p252904,Pintilie10p114111,Cao10p102104,Yang10p143,Ji10p1763,Alexe11p256}.  In particular, bismuth ferrite (BFO) has been found to generate significant bulk photocurrents; combined with its unusually low band gap of 2.7~eV, it has attracted a great deal of attention for its potential in photovoltaic applications~\cite{Choi09p63,Yang10p143,Alexe11p256,Nechache11p202902,Katiyar14p142902,Fang14p142903,Nakashima14p09PA16,Gupta15p062902,Gao15p176,Nechache14p61}. 
The understanding of the fundamental physics behind the effect has advanced as well; 
%Previously, we performed first-principles calculations of the so-called ``shift current'', reproducing experimental results and providing strong evidence that shift currents are the primary mechanism for the bulk photovoltaic effect~\cite{Young12p116601,Young12p236601}.  
recently we demonstrated that the BPVE can be attributed to ``shift currents'', and that the bulk photocurrents may be calculated from first-principles. The {\em ab initio} calculation of the shift current and subsequent analysis yielded several chemical and structural criteria for optimizing the response. These criteria have been used previously to modify or identify existing materials with enhanced response~\cite{Jiang14p075153,Brehm14p204704,Wang15p165124,Zheng15p31}. In this work, we use these insights to propose several candidate bulk photovoltaics with calculated response as much as an order of magnitude higher than well-known ferroelectrics, while having band gaps in or slightly below the visible spectrum. Our results demonstrate that bulk photovoltaic response can be much stronger than previously observed, supporting the possibility of materials suitable for application.

There are two figures of merit for evaluating the BPVE in a material: the current density response to a spatially uniform electric field, and the Glass coefficient~\cite{Glass74p233}.  The current density response is given by the tensor
\begin{flalign*}
J_q(\omega)&=\sigma_{rsq}(\omega) E^0_r(\omega) {E}^0_s(\omega)\\
\sigma_{rsq}(\omega)&=e\sum_{n',n''}\int \dx{\mathbf{k}}\mathcal{I}_{rs}(n',n'',\mathbf{k}; \omega)\mathcal{R}_q(n',n'',\mathbf{k})
\end{flalign*}
where $\mathbf{E}^0$ is the vector of the illumination field, and $n'$ and $n''$ index bands. Letting $f$ denote filling, $\chi$ the Berry connection, and $\phi$ the phase of the transition dipole, the expression 
\begin{flalign}
\mathcal{I}_{rs}(n',n'',\mathbf{k}; \omega)=&\pi \left(\frac{e}{m\hbar\omega}\right)^2\left(f[n'' \mathbf{k}] -f[n' \mathbf{k}]\right)\nonumber\\
&\times\bra{n' \mathbf{k}} \hat{P}_r  \ket{n'' \mathbf{k}} \bra{n'' \mathbf{k}} \hat{P}_s\ket{n' \mathbf{k}}\nonumber\\ 
&\times  \delta \left(\omega_{n''}(\mathbf{k})-\omega_{n'}(\mathbf{k}) \pm \omega\right)
\end{flalign}
describes the intensity of transitions, and  
\begin{flalign}
 \mathcal{R}_q(n',n'',\mathbf{k})=-\dpar{\phi_{n' n''}(\mathbf{k}, \mathbf{k})}{k_q} - \left[\chi_{n''q}(\mathbf{k})-\chi_{n'q}(\mathbf{k})\right]
\end{flalign}
is the expression for the ``shift vector'', which describes a distance associated with the excited carrier~\cite{vonBaltz81p5590,Sipe00p5337}, and depends on the differences in the wavefunction centers up to a unit cell, as provided by the Berry connections, and the average separation in unit cells given by the transition dipole phase derivative.  Roughly speaking, the two terms $\mathcal{I}$ and $\mathcal{R}$ can be thought of as giving the number of carriers excited and the velocity of those carriers.  We emphasize that this mechanism is profoundly different from other photovoltaic effects; rather than relying on excitation of carriers which are then separated by an electric field, the carriers are electron/hole pairs in coherent excited states that possess intrinsic momentum of opposite sign.  Crucially, this allows for arbitrarily high photovoltages; the Schockley-Queisser limit does not apply, a major advantage of BPVE. In particular, the open-circuit voltage is determined by the competition between the photocurrent and countervailing voltage-driven leakage that depends on the overall resistance of the sample~\cite{Bhatnagar13p2835}.  This sample dependence prevents straightforward calculation.  

Determining the total current in a sample is complicated by the attenuation of incident illumination as it travels through the material.  In the limit of a thick sample that will completely absorb the illumination, the total current can be obtained from the Glass coefficient $G$ 
\begin{flalign}
\bar{J}_q(\omega)=\frac{\sigma_{rrq}(\omega)}{\alpha_{rr(\omega)}}\abs{E^0_r(\omega)}^2\mathcal{W}=G_{rrq}(\omega) I_r(\omega)\mathcal{W}
\end{flalign}
where $\alpha$ is the absorption coefficient, and $\mathcal{W}$ is the sample width. Thus, the current density tensor and Glass coefficient describe the response in the regimes of near-zero and near-infinite thickness, respectively. In practice, ``infinite thickness'' is on the order of microns, and total photocurrent is usually best described by the Glass coefficient.

However, the Glass coefficient provides additional information about the response.  In the limit where $\epsilon^i \ll\epsilon^r$, 
\begin{flalign*}
\alpha \approx \frac{\omega}{cn}\epsilon^i=\left(\frac{e}{m}\right)^2\frac{\pi}{\epsilon_0cn\hbar\omega}\sum_{n',n''}\int \dx{\mathbf{k}}\mathcal{I}_{rs}(n',n'',\mathbf{k}; \omega)
\end{flalign*}

and the Glass coefficient becomes
\begin{flalign*}
G_{rrq}(\omega)  =&\frac{1}{2\epsilon_0 cn} \frac{\sigma_{rrq}(\omega)}{\alpha_{rr(\omega)}}\\
                 =&\frac{e}{2\hbar\omega} \frac{\sum_{n',n''}\int \dx{\mathbf{k}}\mathcal{I} _{rr}(n',n'',\mathbf{k}; \omega) R_q(n',n'',\mathbf{k})}{\sum_{n'n''}\int \dx{\mathbf{k}}\mathcal{I}_{rr}(n',n'',\mathbf{k}; \omega)}
\end{flalign*}
The Glass coefficient is therefore closely related to the weighted average shift vector, allowing us to estimate the contribution of both terms in the shift current expression.

Shift current response was calculated as in Ref.~\cite{Young12p116601}, from wavefunctions generated using density functional theory, with the generalized gradient approximation (GGA) and optimized, norm conserving pseudopotentials~\cite{Rappe90p1227, Ramer99p12471}. The presented results exclude spin-orbit effects; calculations with and without spin-orbit were performed for both LiBiO$_3$  Mg$_{1/2}$Zn$_{1/2}$PbO$_3$ and were not found to substantially influence the results.  For BFO, a Hubbard $U$ of 5~eV was used for Fe~$3d$, as in Ref.~\cite{Young12p236601}. For PbNiO$_3$, a Hubbard $U$ of 4.6~eV was used for Ni~$3d$, as in Ref.~\cite{Hao12p014116}. QUANTUM ESPRESSO~\cite{Giannozzi09p395502} was used for the electronic structure calculations, and OPIUM was used to generate pseudopotentials.  The Heyd-Scuseria-Ernzerhof (HSE) hybrid functional~\cite{Heyd03p8207} was used to compute band gaps, as it is known to frequently produce significantly more accurate values than GGA. These calculations were performed on $8\times 8\times 8$ k-point grids, with $4\times 4\times 4$ grids for the exact exchange HSE calculations.  Band structures and density of states plots are generated from GGA calculations, and reported HSE band gaps are the direct gaps.
The present results are for the experimental structure in the cases of LNO, BFO, and PbNiO$_3$, and computationally relaxed structures for the other materials. Structural relaxations and calculations of the shift current were performed at the level of LDA and found to vary minimally from the GGA results; due to the high expense of exact exchange calculations, the dense k-point grids required to converge shift current calculations cannot presently be obtained using HSE, and scissor corrections~\cite{Levine89p1719,Nastos05p045223} to the HSE gaps were applied to account for the dependence of Glass coefficient on frequency.

Previously, we revealed the dependence of shift vector magnitude on the chemical and structural properties of materials.  Large shift vectors were characterized by valence and/or conduction states that are both strongly asymmetric and delocalized in the current direction~\cite{Young12p116601}.
In this regard, many distorted perovskite ($AB$O$_3$) ferroelectrics are crippled by the presence of $d^0$ cations enclosed in octahedral oxygen cages.  The conduction band edge is dominated by $t_{2g}$-like $d$ states that are largely nonbonding.  Coupled with the tendency for $d$ states to localize, the result is that both shift vectors and transition response are very weak near the band gap. The delocalized $e_g$ states are much higher in energy, effectively raising the energy threshold for significant BPVE.
\begin{figure}
{
\includegraphics[width=3.0in]{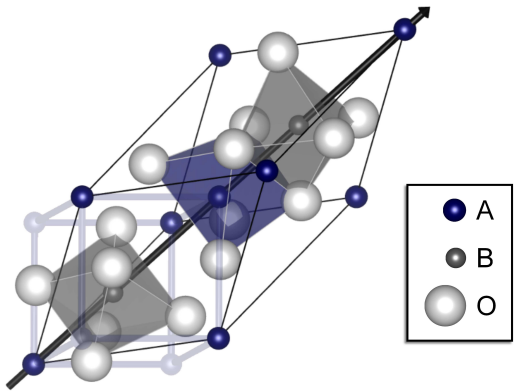}
}

\caption{The LiNbO$_3$ (LNO) primitive cell, overlaid with the pseudo-cubic perovskite cell.  The direction of the polar distortion is indicated by the black arrow.}\label{fig:struct}
\end{figure}
\begin{table*}
\begin{tabular}{l | c | c | c}
\hline
         & PbNiO$_3$ & Mg$_{1/2}$Zn$_{1/2}$PbO$_3$ & LiBiO$_3$\\ 
\hline
a        & 5.63~\AA & 5.77~\AA  & 5.67~\AA \\ 
$\alpha$ & 57$^\circ$       &  57$^\circ$ & 56$^\circ$\\
$A$~(2a) &  (0.0, 0.0, 0.0) & (0.0, 0.0, 0.0) & (0.0, 0.0, 0.0) \\ 
$B$~(2a) & (0.214, 0.214, 0.214) & (0.216, 0.216, 0.216)& (0.213, 0.213, 0.213)\\ 
O~(6b)   & (0.830, 0.098, 0.415)  & (0.794, 0.128, 0.390)& (0.798, 0.122, 0.405)\\
P        &  99~$\mu$C/cm$^2$  &   83~$\mu$C/cm$^2$  & 50~$\mu$C/cm$^2$\\
\hline
\end{tabular}
\caption{The structural data for the three compounds presented here. PbNiO$_3$ and  LiBiO$_3$ are in space group R3c, while  Mg$_{1/2}$Zn$_{1/2}$PbO$_3$ is in R3.  However, the deviations of the coordinates from the R3c positions are miniscule $(<0.2\%)$, so they are presented as such with Mg and Zn each occupying one site of the $A$ position. Polarizations were determined based on a non-polar structure featuring the $A$-site atom coplanar with oxygen, and the $B$-site midway between oxygen planes~\cite{Rabe07p1,Neaton05p014113,Levchenko08p256101,Baeumer15p6136}. }
\label{tab:struct}
\end{table*}

\begin{figure}
{
\subfigure[]{ \includegraphics [width=1.6in]{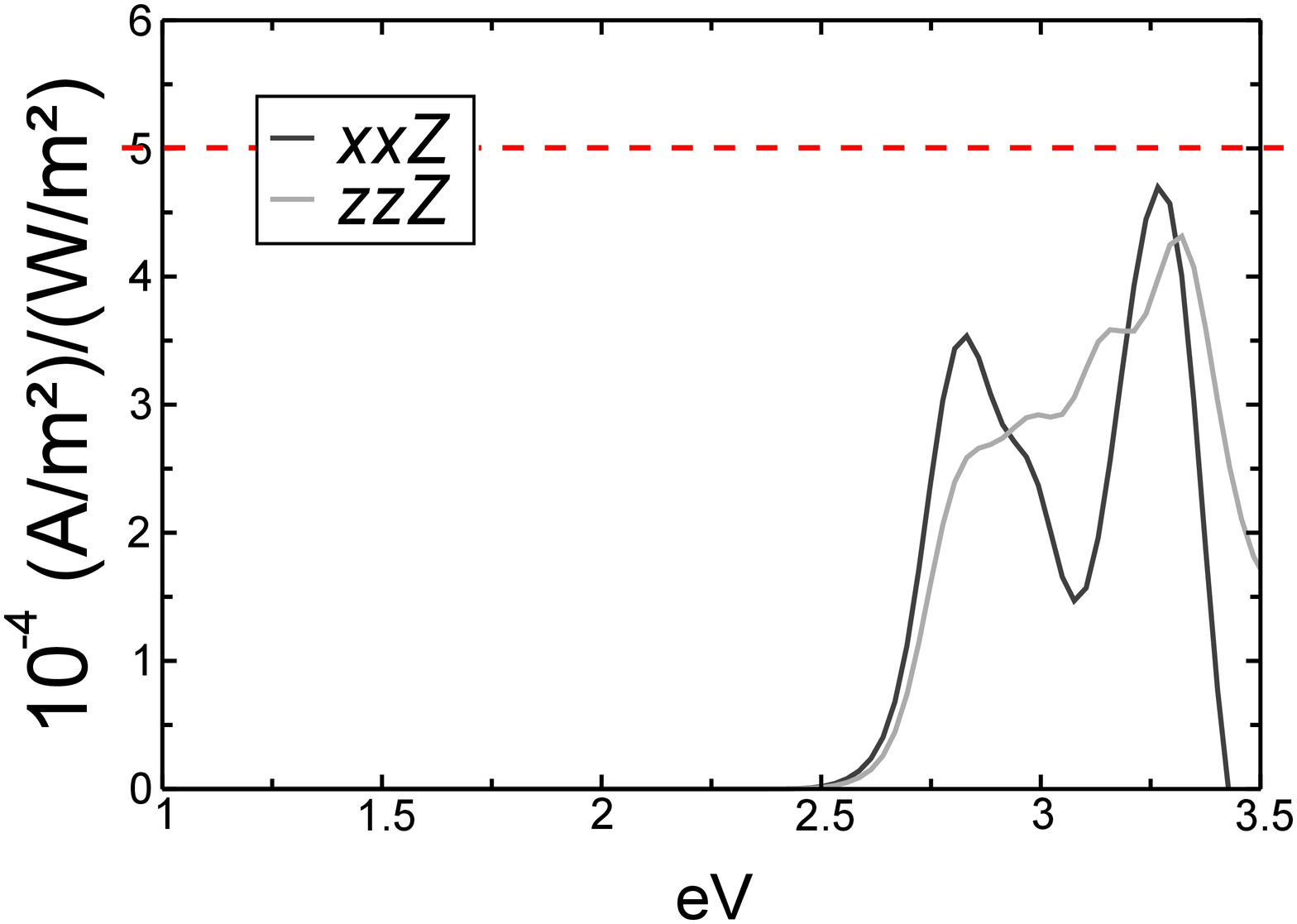}\label{fig:scbfo}}
\subfigure[]{ \includegraphics [width=1.6in]{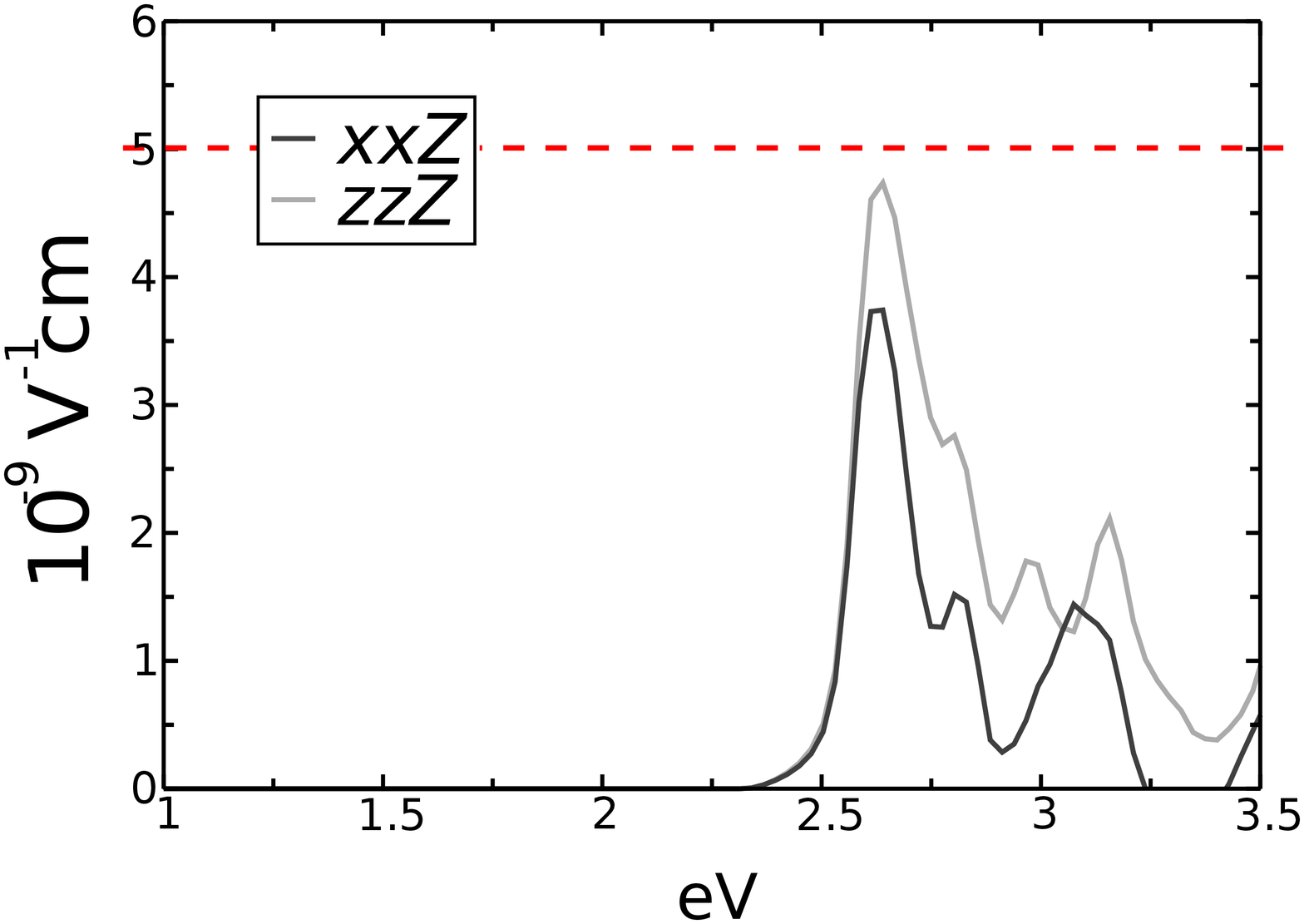}\label{fig:glassbfo}}
}

\caption{The current density response for BFO (GGA+$U$) is shown in~\subref{fig:scbfo}. The Glass coefficient of BFO appears in~\subref{fig:glassbfo}.  Only the response in the direction of material polarization is shown, for both perpendicular ($xxZ$) and parallel ($zzZ$) light polarization. Dashed lines appear at benchmark values of current density and Glass coefficient chosen to represent the maximum response of these materials. }\label{fig:lnobfo}
\end{figure}

To overcome the weak BPVE response of $d^0$ oxides, we investigated systems that involve both large distortions to oxygen cages, (increasing the bonding character of any $d^0$ states) as well as $d^{10}$ cations with less localized $s$ and/or $p$ states near the band edge~\cite{Jiang14p075153}.  It has already been noted that $d^{10}$ cations can dramatically improve the activity of photocatalysts~\cite{Inoue09p364}. We found polar oxides taking the LiNbO$_3$ structure to be promising candidates, with $d^{10}s^0$ cations Pb$^{4+}$ and Bi$^{5+}$. This structure can also be obtained by distorting the perovskite structure rhombohedrally, and allowing polar distortions along and oxygen-cage rotations about $\left<111\right>$.  Notable ferroelectrics with this structure (but with $d^0$ cations) include LiNbO$_3$ (LNO) and BiFeO$_3$ (BFO).  LNO is known for its large nonlinear optical response, and, often doped with iron, it was one of the first materials in which the bulk photovoltaic effect was observed and studied~\cite{Chen69p3389, Reznik85p215, Anikiev85p89}. However, its bulk band gap is well outside the visible spectrum~\cite{Dhar90p5804}.   BFO has garnered much attention recently for its multiferroic behavior~\cite{Catalan09p2463} and low band gap of about 2.74~eV~\cite{Ihlefeld08p142908}, which has led to explorations of its photovoltaic response~\cite{Choi09p63,Yang10p143,Seidel11p126805,Ji10p1763, Ji11p094115,Alexe11p256}.  We have used BFO as a benchmars for the present study; as with the archetypal ferroelectrics BaTiO$_3$ and PbTiO$_3$, its LUMO is dominated by cation $d$-states and yields a very similar response magnitude.

We consider only current response in the direction of material polarization for both perpendicular ($xxZ$) and parallel ($zzZ$) light polarization, as these are the only tensor elements that can contribute to the response to unpolarized light.  For ease of comparison, we mark baseline values reflecting the maximum response of our benchmark, shown in Fig.~\ref{fig:lnobfo}, with a dashed, red line.  These are, for the current density and Glass coefficient, respectively, $5\times 10^{-4}({\rm A/m^2})/({\rm W/m^2})$ and $5\times 10^{-9}$cm/V.

\begin{figure*}
{
\subfigure[]{ \includegraphics [width=2.1in]{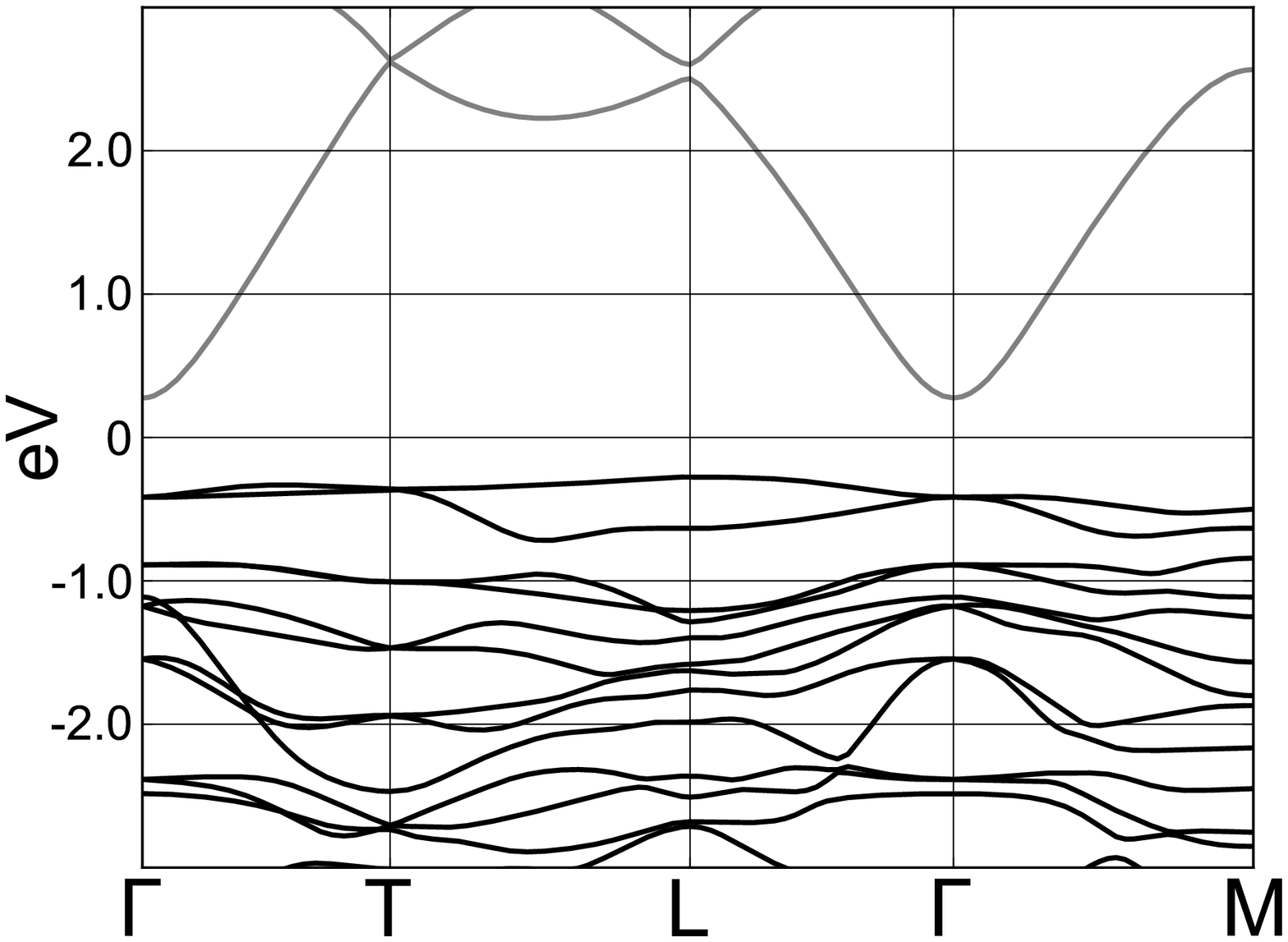}\label{fig:bandsnipb}}
\subfigure[]{ \includegraphics [width=2.1in]{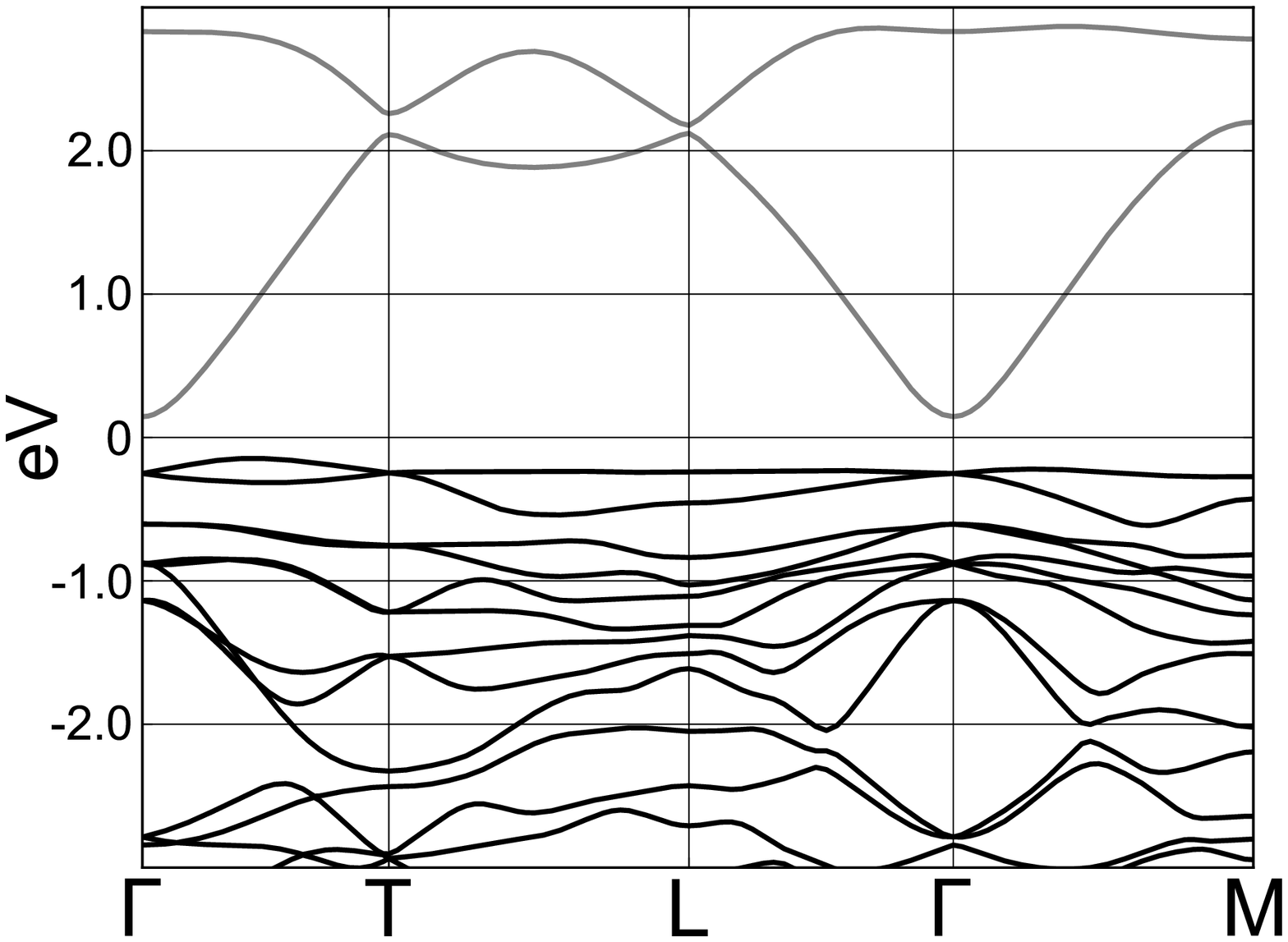}\label{fig:bandsmzpb}}
\subfigure[]{ \includegraphics [width=2.1in]{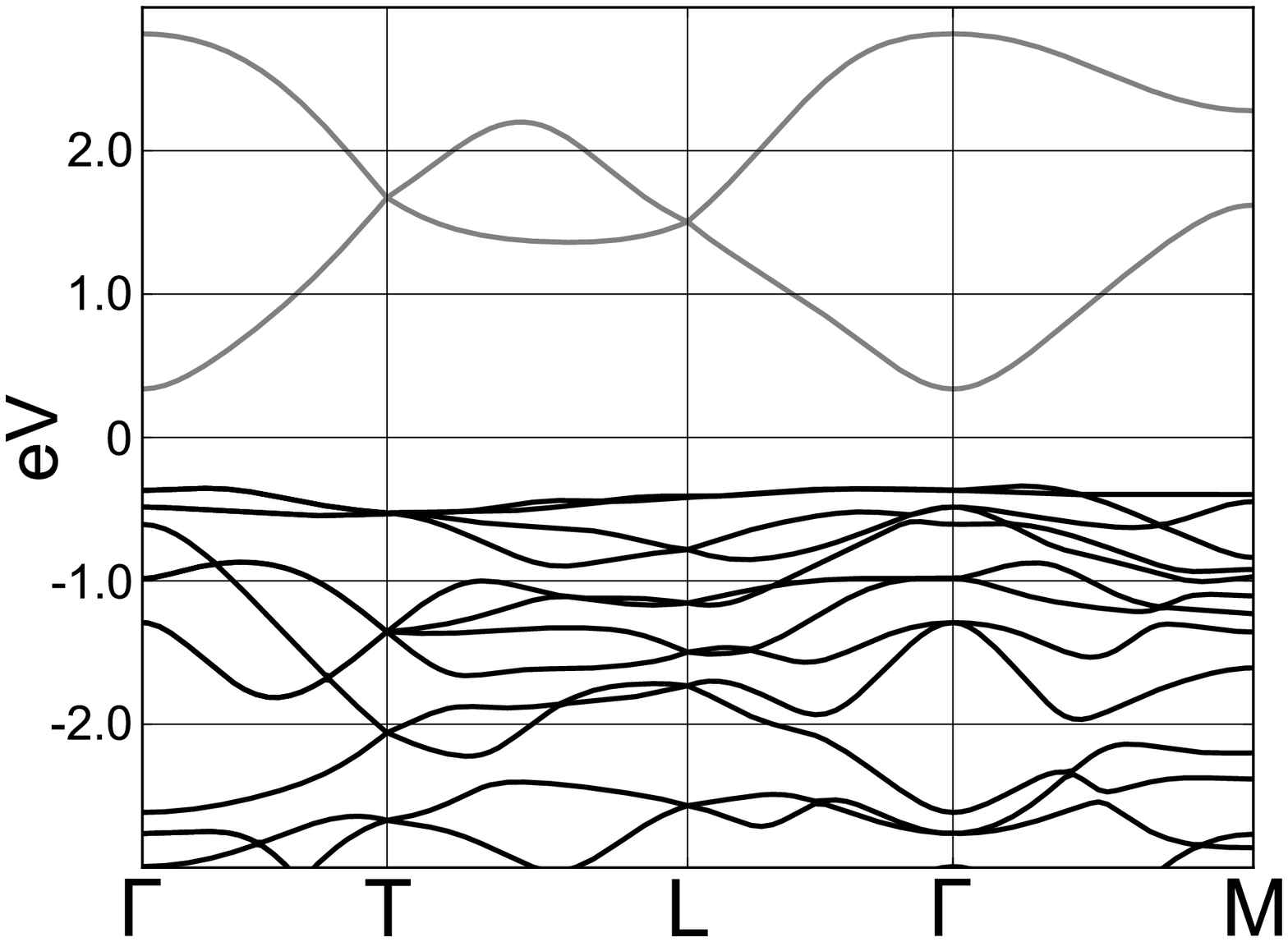}\label{fig:bandslibi}}
}
\caption{The band structures of~\subref{fig:bandsnipb} PbNiO$_3$, ~\subref{fig:bandsmzpb} Mg$_{1/2}$Zn$_{1/2}$PbO$_3$, and~\subref{fig:bandslibi} LiBiO$_3$. Note the similar, highly dispersive conduction band edges.}\label{fig:bands}
\end{figure*}

We have studied three materials taking the LNO structure (Fig~\ref{fig:struct}): PbNiO$_3$, Mg$_{1/2}$Zn$_{1/2}$PbO$_3$, and LiBiO$_3$.  The first has been synthesized~\cite{Inaguma11p16920}, and the latter two are similar in composition to known materials.  The structural parameters and bulk polarizations are given in Table~\ref{tab:struct}. The distortion from cubic perovskite is sufficiently strong that assignment of $A$- and $B$-sites is ambiguous; we have followed the assignment of Ref.~\cite{Inaguma11p16920} for PbNiO$_3$, but note that treating Ni as the $A$-site (reversing the orientation), the Wyckoff position of oxygen becomes $(0.800,0.116,0.384)$, which is slightly closer to the corresponding crystal coordinate of the other two materials.   All three satisfy our requirements of low band gap, $d^{10}$ cations, and large polar distortions. Furthermore, as seen in Fig.~\ref{fig:bands}, all three have qualitatively similar band structures, featuring highly dispersive conduction bands, in contrast to the usual case of $d^0$ perovskite derivatives.  As we will show, this arises due to unfilled $s$-like -- rather than $d$-like -- states composing the conduction band, and has profound consequences for the bulk photovoltaic response.

\begin{figure}
{
\subfigure[]{ \includegraphics [width=1.6in]{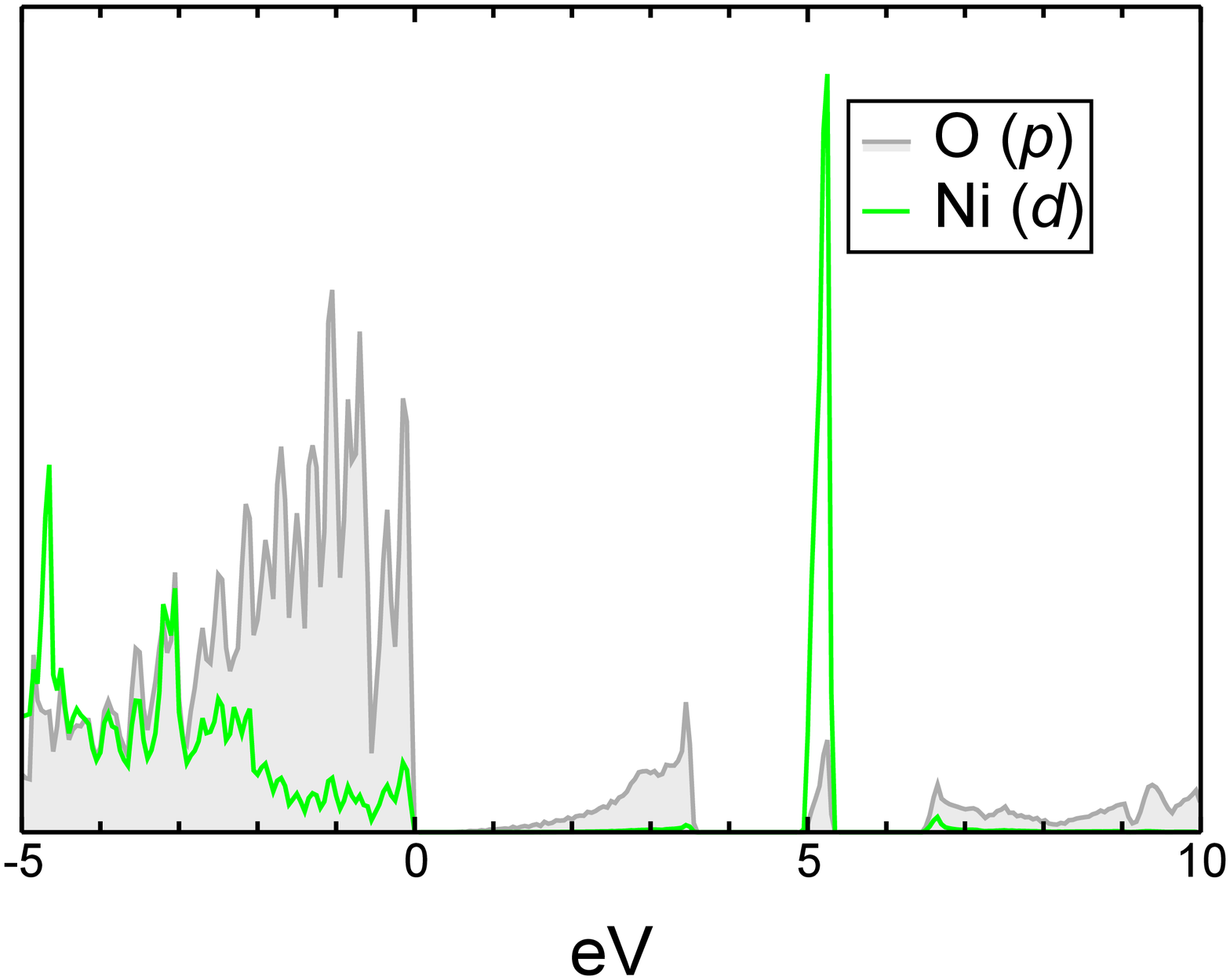}\label{fig:pdosnipbA}}
\subfigure[]{ \includegraphics [width=1.6in]{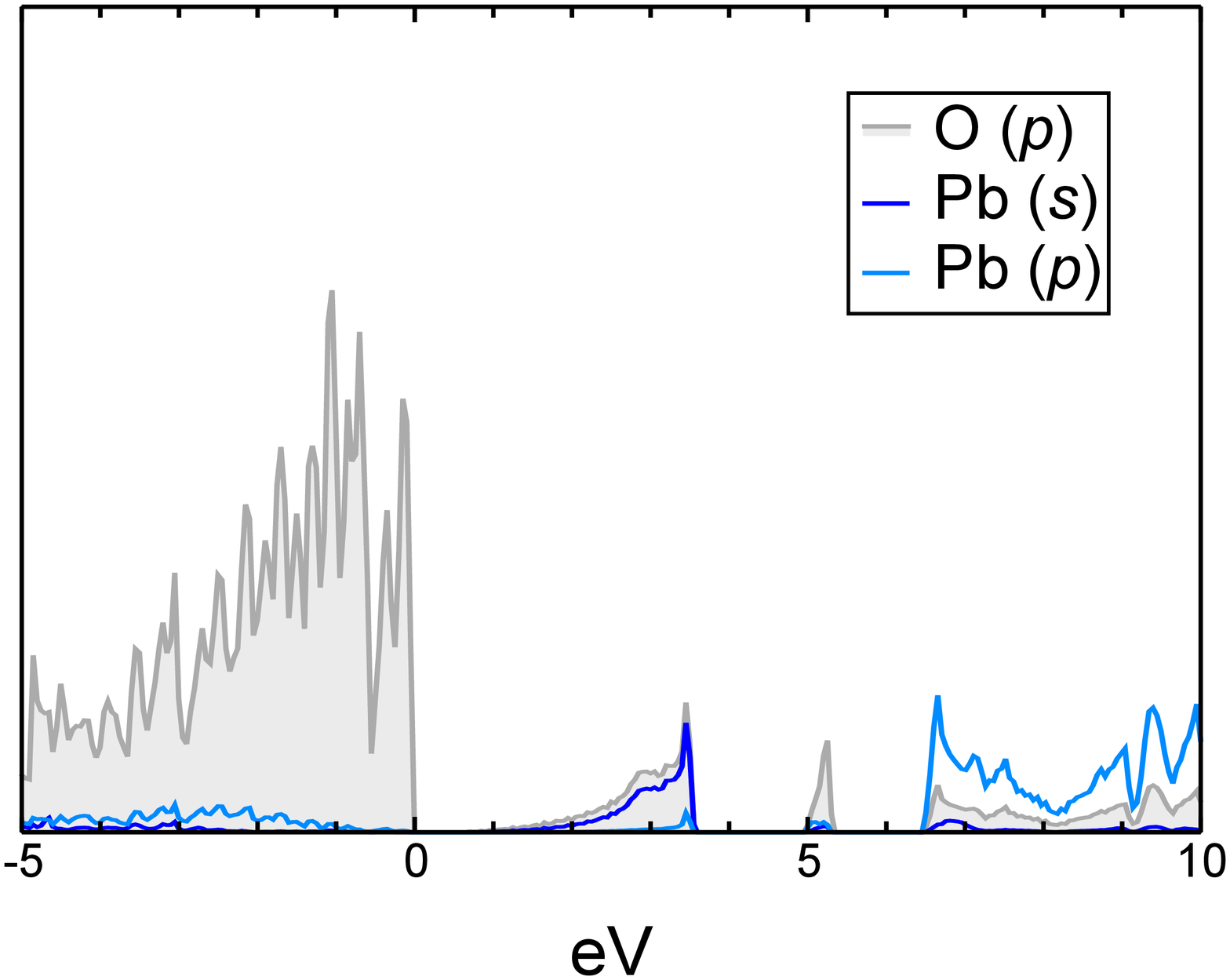}\label{fig:pdosnipbB}}
\subfigure[]{ \includegraphics [width=1.6in]{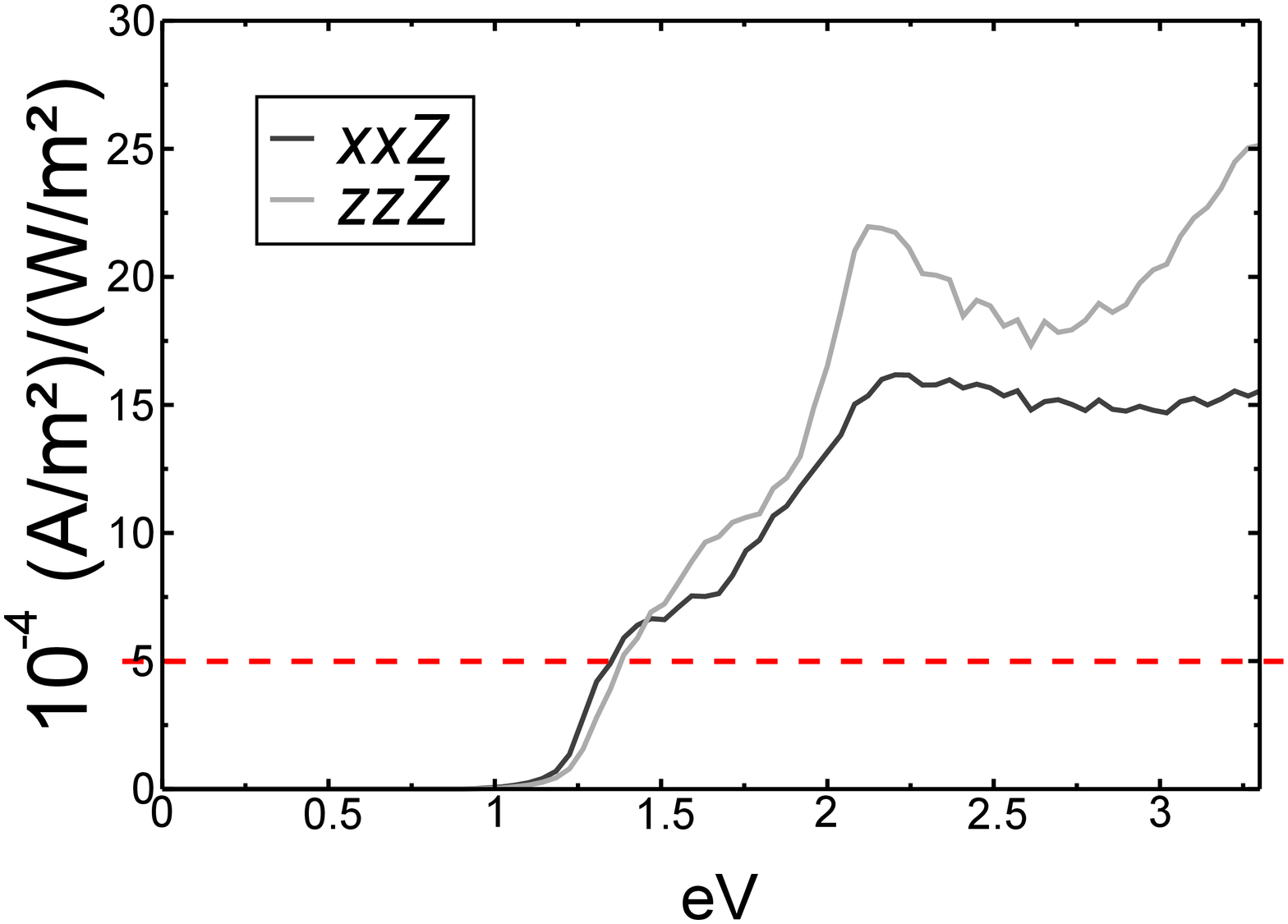}\label{fig:scnipb}}
\subfigure[]{ \includegraphics [width=1.6in]{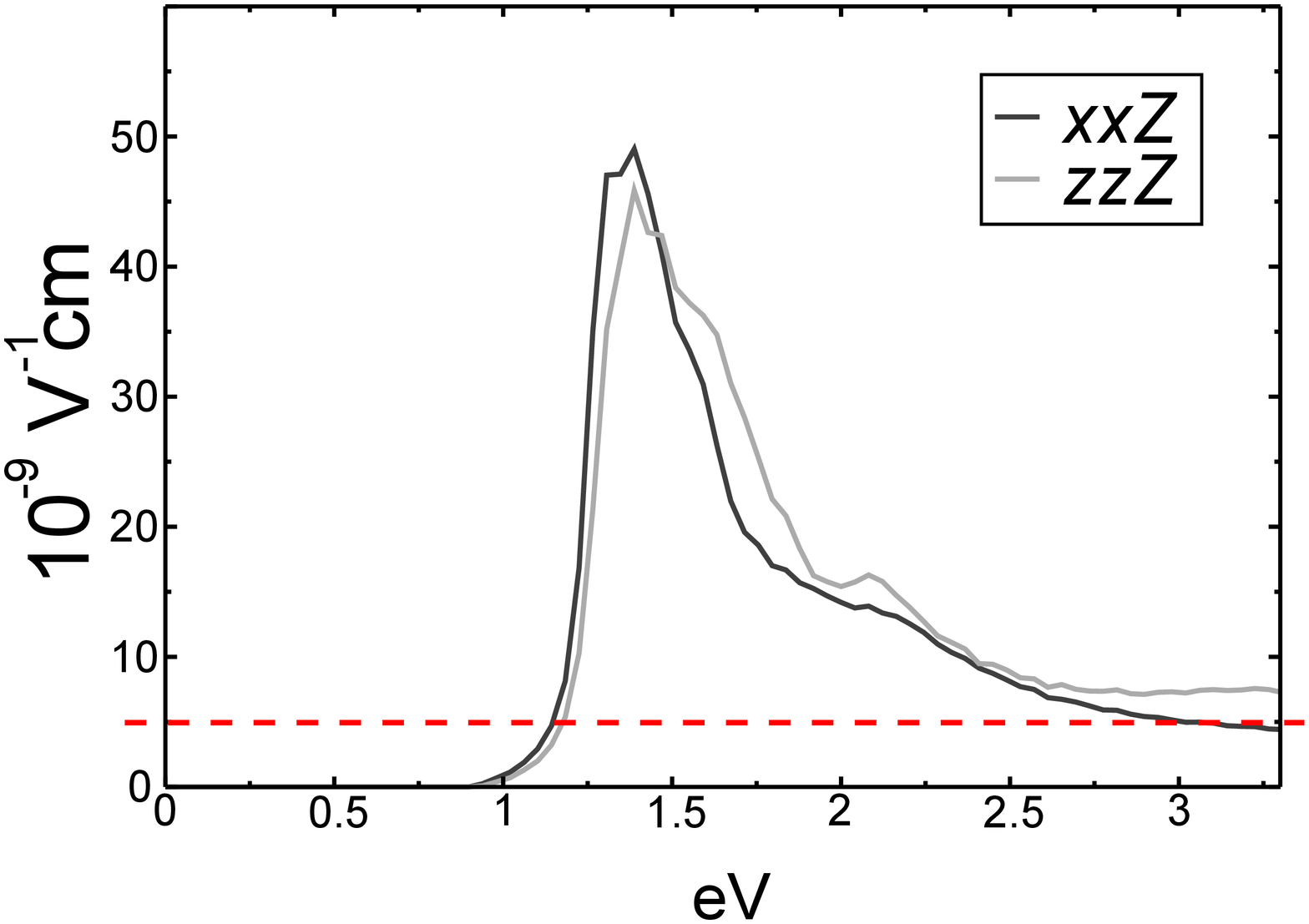}\label{fig:glassnipb}}
}
\caption{\subref{fig:pdosnipbA} and~\subref{fig:pdosnipbB} give the projected density of states for PbNiO$_3$. The unfilled half of $e_g$ of the high-spin $d^8$ Ni appears as a sharp peak {\em above} the unfilled Pb $s$-orbitals, which have strongly hybridized with oxygen $p$-orbitals, resulting in a low band gap (1.2eV in HSE~\cite{Hao12p014116}).  This material has a large~\subref{fig:scnipb} current density response ($\approx 4\times$ benchmark), and a very large~\subref{fig:glassnipb} Glass coefficient($\approx 10\times$ benchmark).  }\label{fig:nipb}
\end{figure}

PbNiO$_3$ has recently been synthesized~\cite{Inaguma11p16920} and explored theoretically~\cite{Hao12p014116,Hao14p015030}. Like BFO, it is antiferromagnetic with weak spin-canting, and possesses an even larger polarization, calculated at 100~$\mu$C/cm$^2$~\cite{Hao12p014116}.  Its band gap is even lower than BFO, with HSE predicting 1.2~eV~\cite{Hao12p014116}. In BFO, Bi has oxidation state $3+$, so that its $6s$ orbital is filled, and the exchange splitting of Fe determines the gap.  However, in PbNiO$_3$, Pb is $4+$, and its $6s$-states appear lower in energy than the Ni exchange-split bands, resulting in a distinct electronic profile.  This can be clearly seen in the projected density of states (Fig.~\ref{fig:pdosnipbA}): the lowest conduction band is almost entirely Pb~$6s$ and O~$2p$ states, while the $d$-states only appear in the valence band and higher in the conduction manifold. While this serves to lower the band gap dramatically, a further result of this is a Glass coefficient (Fig.~\ref{fig:glassnipb}) over an order of magnitude larger than the benchmark value.  The current density is modest by comparison, though it still exceeds the benchmark, indicating large shift vectors with relatively low absorption.

\begin{figure}
{
\subfigure[]{ \includegraphics [width=1.6in]{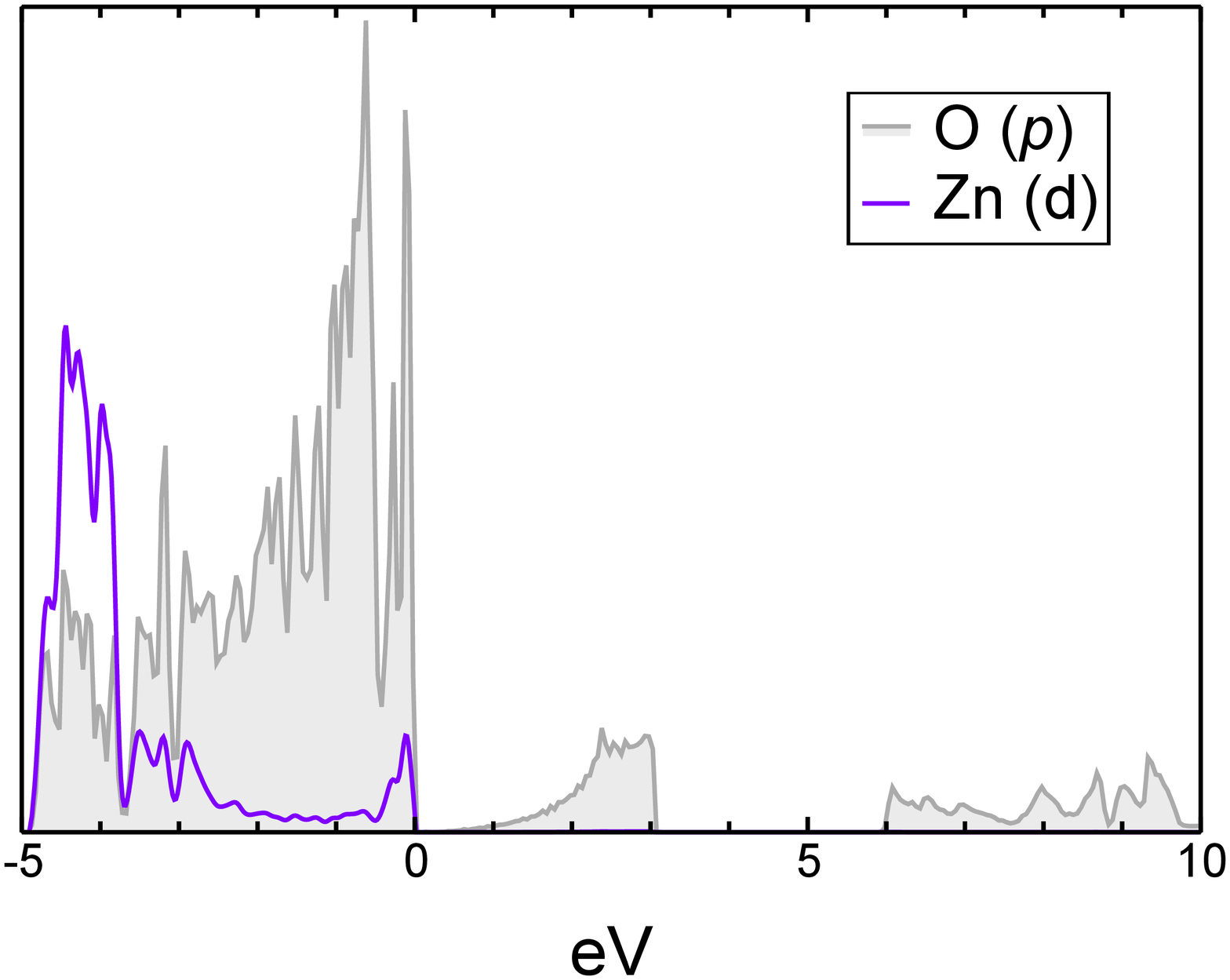}\label{fig:pdosmzpbA}}
\subfigure[]{ \includegraphics [width=1.6in]{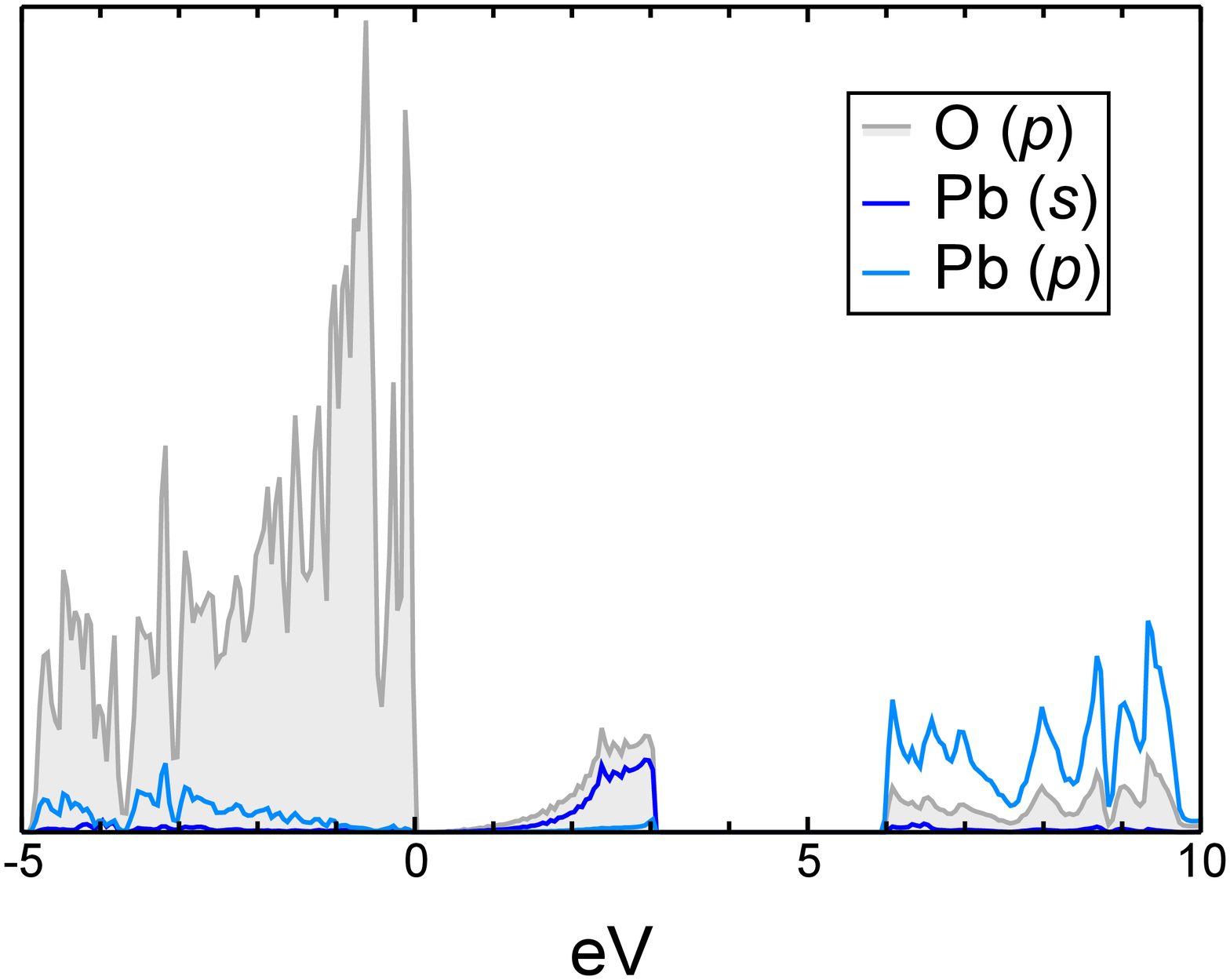}\label{fig:pdosmzpbB}}
\subfigure[]{     \includegraphics [width=1.6in]{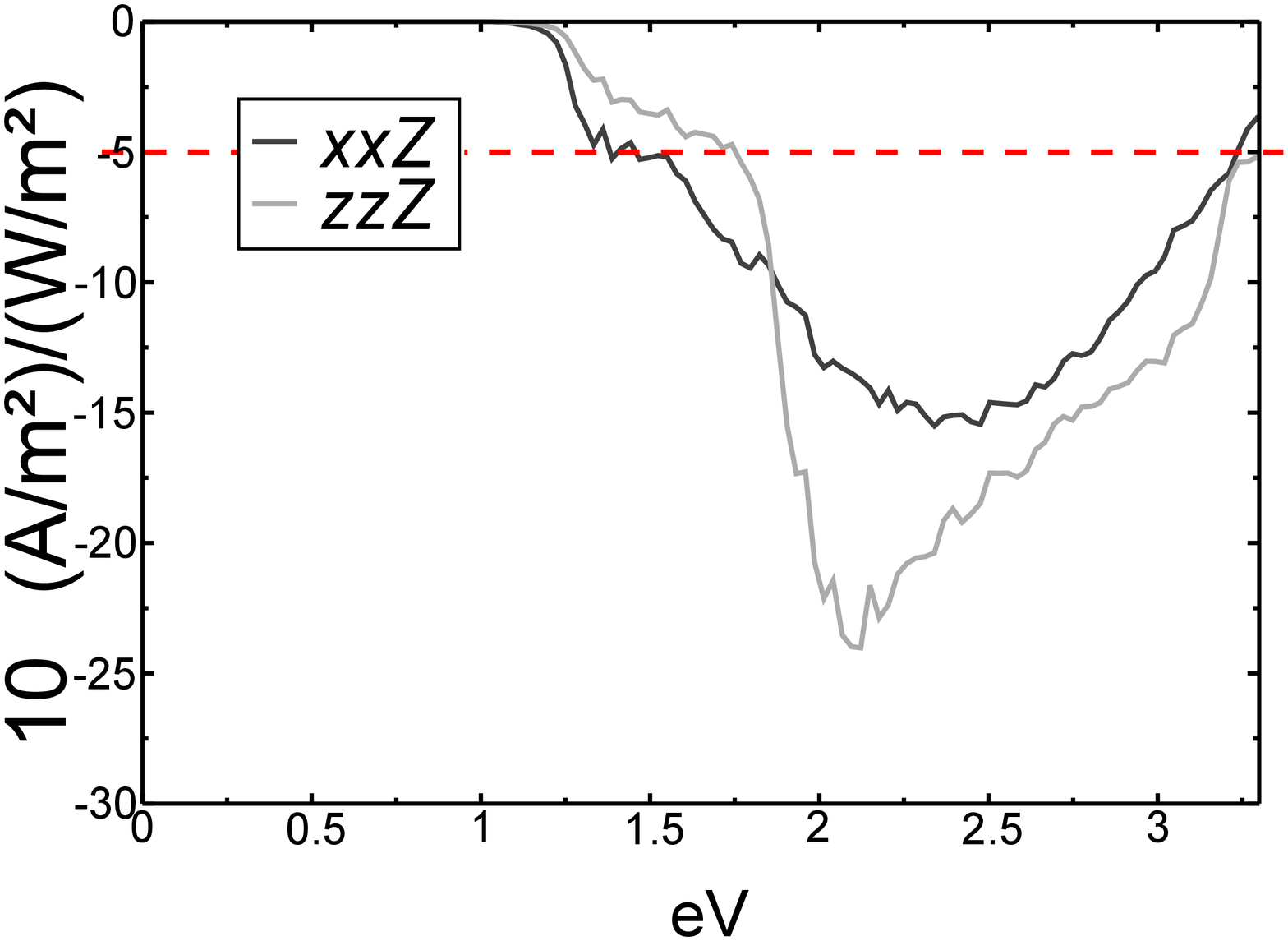}\label{fig:scmzpb}}
\subfigure[]{     \includegraphics  [width=1.6in]{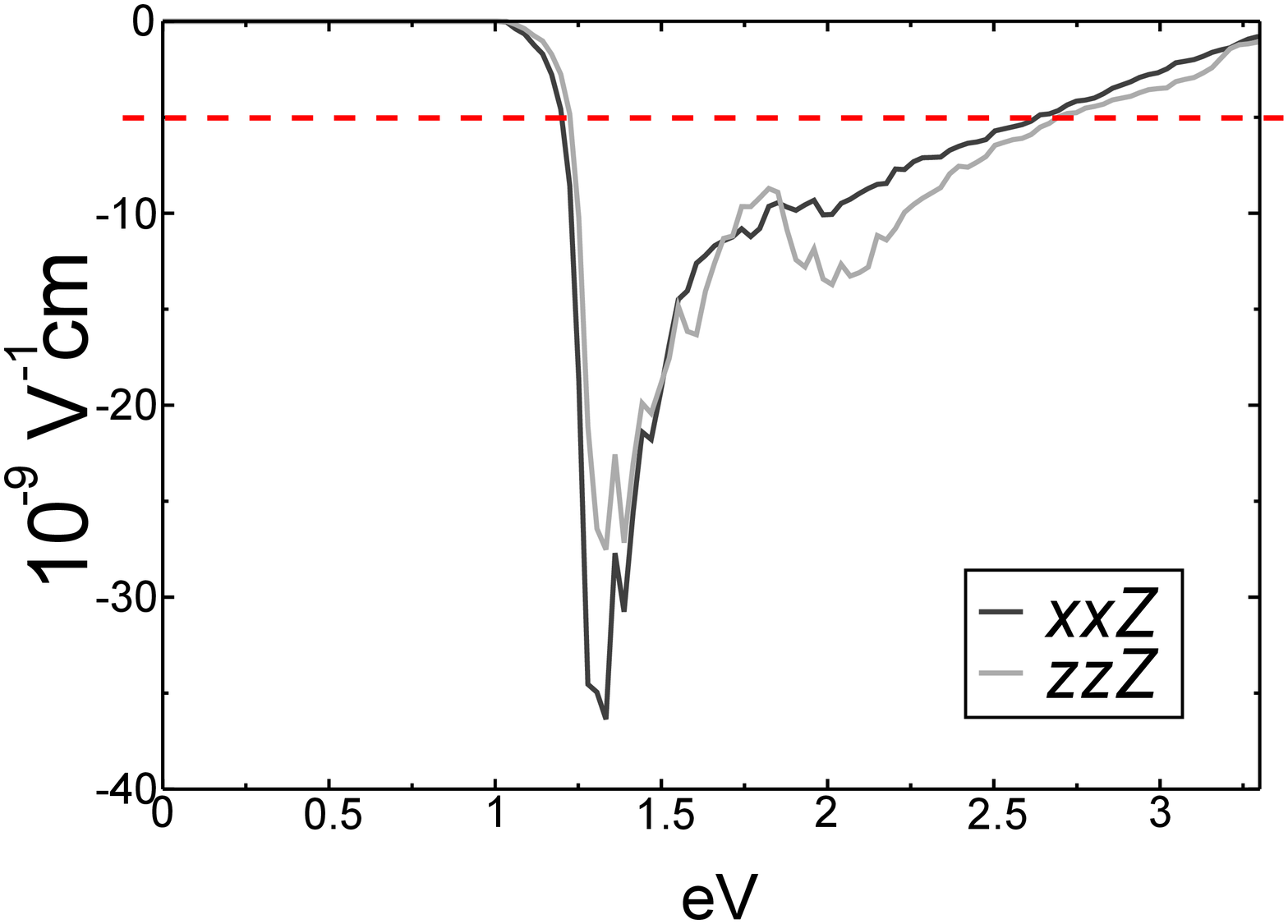}\label{fig:glassmzpb}}
}
\caption{Orbital-projected densities of states for Mg$_{1/2}$Zn$_{1/2}$PbO$_3$ are shown in~\subref{fig:pdosmzpbA} and~\subref{fig:pdosmzpbB}.  The valence band is formed almost entirely from oxygen $p$-orbitals, and the conduction band is hybridized Pb~$6s$ and O~$2p$-states.  This results in a low band gap (1.2 in HSE), ~\subref{fig:scmzpb} high current density response ($\approx 4\times$ benchmark), and~\subref{fig:glassmzpb} a very large Glass coefficient ($\approx 7\times$ benchmark). }\label{fig:mzpb}
\end{figure}

HgPbO$_3$~\cite{Sleight73p509} and ZnSnO$_3$~\cite{Inaguma08p6704,Gou10p552} are known to take the ilmenite and LiNbO$_3$ structures, respectively. However, the first is metallic and the second has a high band gap and only modest photovoltaic response. We first calculated the response of ZnPbO$_3$, but found it to be borderline metallic, despite promising response; to raise the gap, we substituted Mg for half of the Zn.  Phonon calculations indicate that the structure is metastable.  Once again, as seen in Fig.~\ref{fig:pdosmzpbB}, hybridized Pb~$6s$ states compose the lowest unfilled band.  The magnitude of the response is quite high, but the current is antiparallel to the computed polarization.  This is unlike most materials, including our benchmark materials and the aforementioned PbNiO$_3$, however, we emphasize the ambiguity of both polarization (the addition of polarization quanta) and structure orientation (designation of $A$ and $B$ cations) for these materials.  If we compare the two Pb compounds with Pb as the B-site in both, not only are the structures more similar, but their responses become parallel.

LiBiO$_3$ is known to exist in a structure with edge sharing oxygen octahedra~\cite{Takei11p2017}.  However, our calculations place the LiNbO$_3$-type structure -- which phonon analysis reveals to be metastable -- nearby in energy, at only about 0.01~eV per atom higher; additionally, NaBiO$_3$ is known to take the closely-related ilmenite structure~\cite{Takei11p2017}.  In light of this, we consider it highly possible that the LiBiO$_3$ can be synthesized in the LiNbO$_3$ structure.

\begin{figure}
{
\subfigure[]{ \includegraphics [width=1.6in]{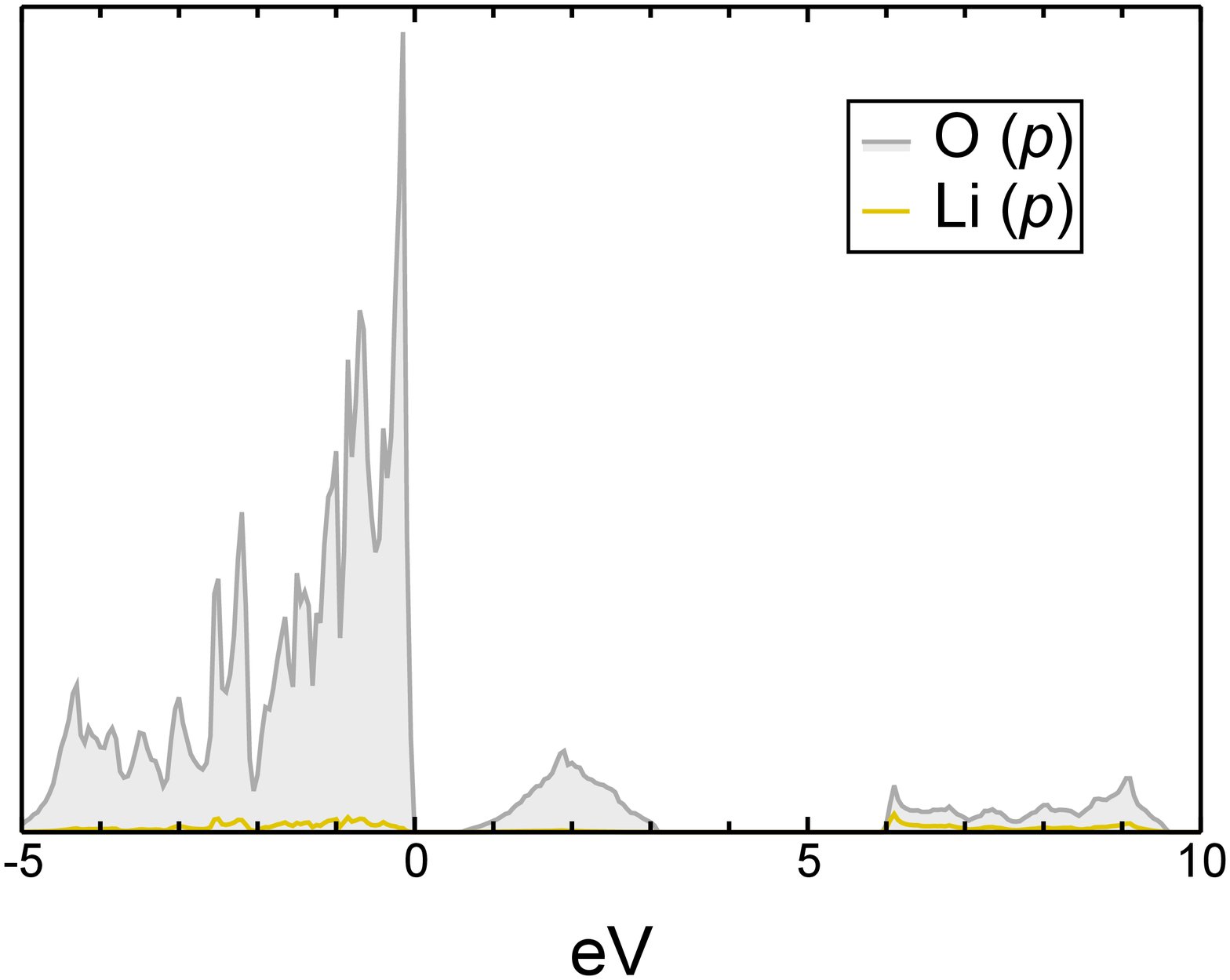}\label{fig:pdoslibiA}}
\subfigure[]{ \includegraphics [width=1.6in]{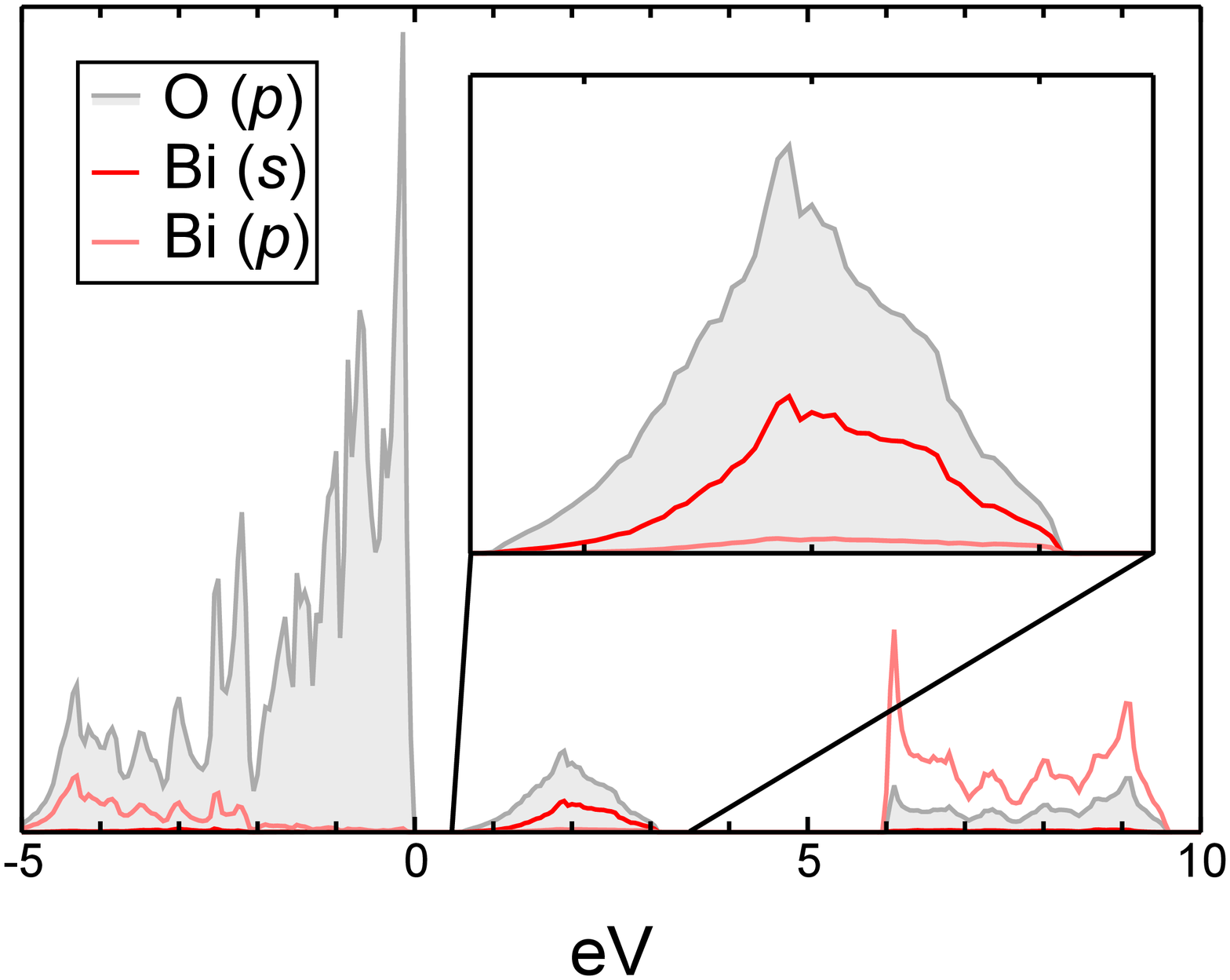}\label{fig:pdoslibiB}}
\subfigure[]{     \includegraphics [width=1.6in]{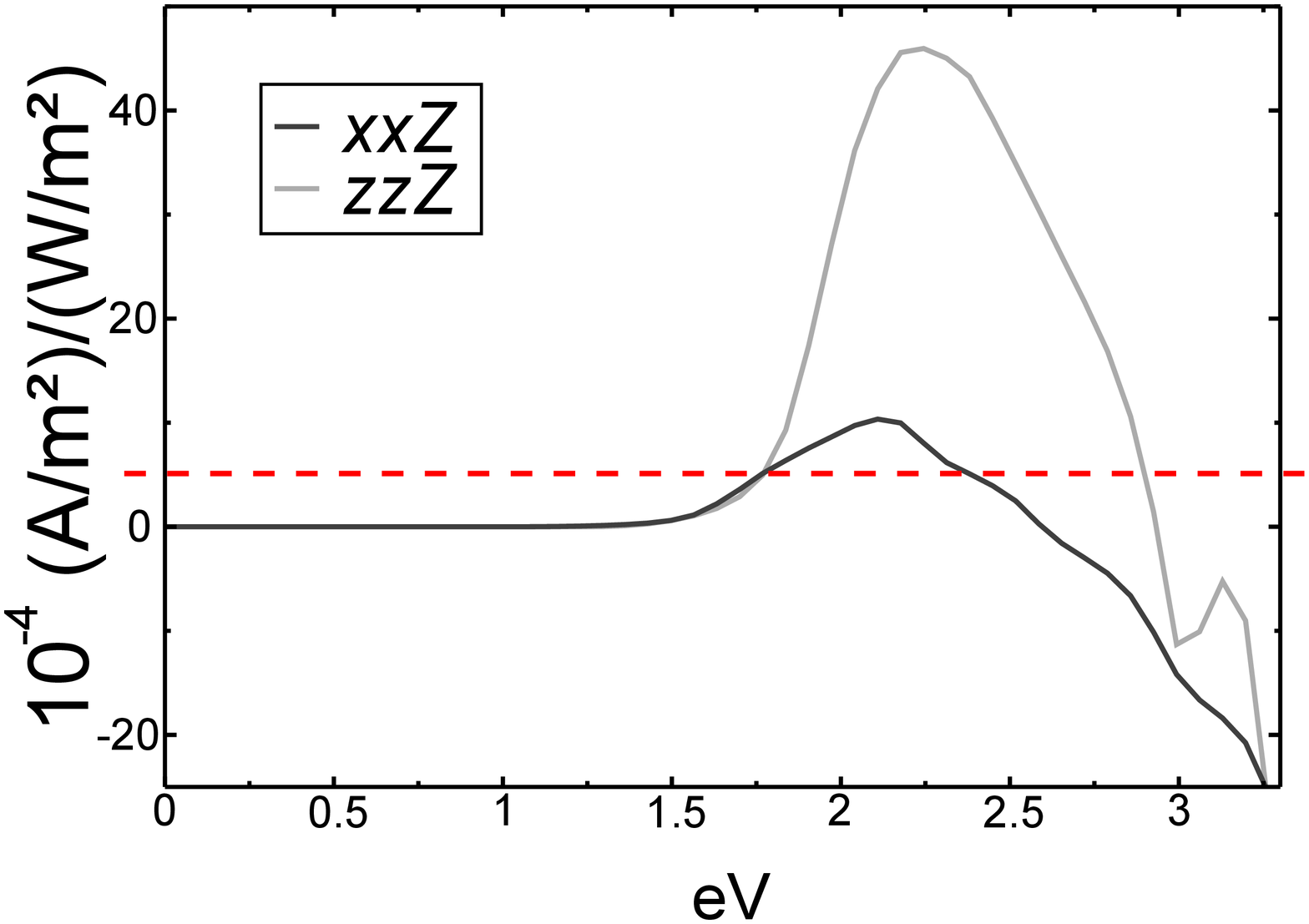}\label{fig:sclibi}}
\subfigure[]{     \includegraphics [width=1.6in]{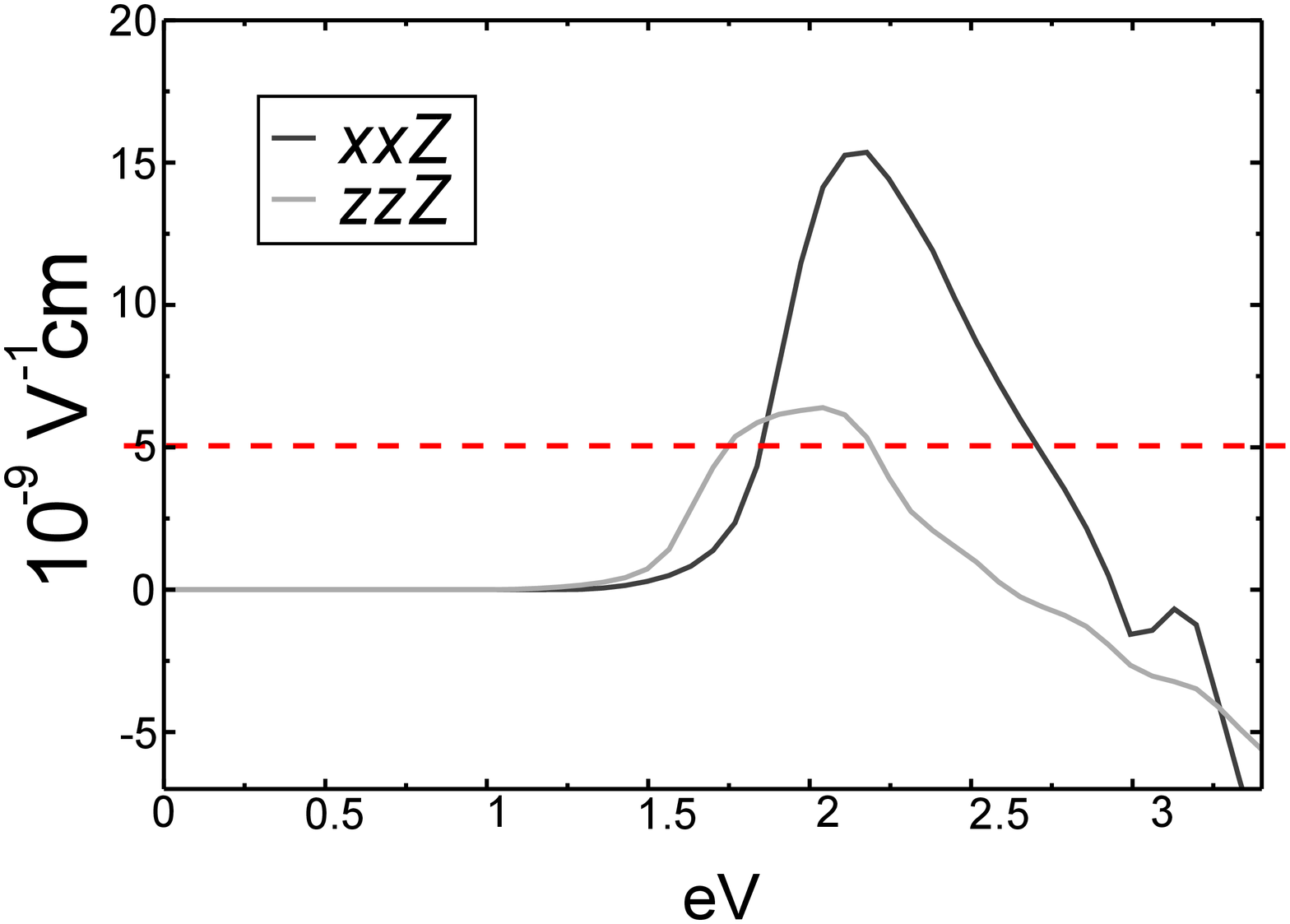}\label{fig:glasslibi}}
}

\caption{ The density of states for LiBiO$_3$, shown in~\subref{fig:pdoslibiA} and~\subref{fig:pdoslibiB}, is dominated by bismuth and oxygen.  The band gap is set by transitions from O~$2p$ to hybridized Bi~$6s$ states. The band gap is modest (1.7eV in HSE).  The current density response, shown in~\subref{fig:sclibi} is quite high ($\approx 8\times$ benchmark), with a large~\subref{fig:glasslibi} Glass coefficient ($\approx 3\times$ benchmark), indicating strong absorption in addition to long shift vectors.  }\label{fig:libi}
\end{figure}

As shown in Fig.~\ref{fig:pdoslibiA} and Fig.~\ref{fig:pdoslibiB}, the electronic structure is very similar to the previous two materials.  As with Pb-containing compounds, the low-lying hybridized Bi~$s$ states form the lowest unfilled bands, though the Bi~$s$ proportion is lower than that of Pb~$s$ in the aforementioned materials. Possibly as a consequence, the dispersion of the conduction band is reduced compared to PbNiO$_3$ and  Mg$_{1/2}$Zn$_{1/2}$PbO$_3$ (Fig.~\ref{fig:bands}), and the BPVE response is somewhat different: while the Glass coefficient is not as large as for the two lead-containing materials, the photocurrent density is higher, indicating increased absorption.  Additionally, the band gap is larger, with HSE predicting $1.7$-$1.8$~eV, positioned almost perfectly with respect to the visible spectrum for solar energy conversion.
\begin{table}
\begin{tabular}{l | l | r | r}
\hline
                              &  \quad Direct gap      & Max. $G$   & Max. $\sigma$    \\ 
                              &  \qquad $\times$ eV  &  $\times 10^{-9}$cm/V   \qquad    &  $\times 10^{-4}\frac{{\rm A/m^2}}{{\rm W/m^2}}$\\
\hline
BiFeO$_3$                     & \quad 2.7~\cite{Ihlefeld08p142908} &   5     \qquad &  5     \qquad\\
\hline
KBNNO~\cite{Wang15p165124}    & \quad 1.3(HSE)                     &  25     \qquad &  5 \qquad \\
LiAsSe$_2$~\cite{Brehm14p204704} & \quad 1.1~\cite{Bera10p3484}   &  98     \qquad & 30 \qquad\\
\hline 
PbNiO$_3$                     & \quad 1.2(HSE)~\cite{Hao12p014116} &  50     \qquad & 25     \qquad \\ 
Mg$_{1/2}$Zn$_{1/2}$PbO$_3$   & \quad 1.2(HSE)                     &  35     \qquad & 25     \qquad \\ 
LiBiO$_3$                     & \quad 1.7(HSE)                     &  15     \qquad & 45     \qquad\\
\hline
\end{tabular}
\caption{The band gap and response characteristics of the presented materials, along with benchmark BiFeO$_3$ and other recently proposed bulk photovoltaics (with scissor corrected responses) for comparison.  }
\label{tab:summary}
\end{table}

It is worth contrasting the response of LiBiO$_3$ with that of the two lead compounds: the former has a notably distinct response, especially near the band edge.  This can be attributed to the difference in valence band character; in PbNiO$_3$ and Mg$_{1/2}$Zn$_{1/2}$PbO$_3$ the valence band edge contains considerable density from the secondary cation filled $d$-states.  This alters the character of the wavefunctions and improves delocalization and response magnitude by sharing density, as opposed to lithium's almost completely ionic character. This suggests that inclusion of an appropriate dopant with higher electronegativity may allow for significant tuning of the response in LiBiO$_3$.

We have proposed several polar oxides in the LiNbO$_3$ structure with strong computed BPVE response and low band gaps, summarized in Table~\ref{tab:summary}.  The compositions, featuring Pb$^{4+}$ or Bi$^{5+}$ cations, were chosen for the absence of $d$-states at the band edge. Instead, these materials have conduction bands formed by low-lying $s$-states hybridized with oxygen $p$-states.  In addition to creating significantly lower band gaps, this makes for large, diffuse orbitals and strongly delocalized states; combined with large polar distortions, they effect significant shift current response that is over an order of magnitude higher than that previously observed, and roughly double the best performing materials previously proposed. Given the minimal contributions from the other cations, the possibility of tuning the response via composition without altering its fundamental character is strongly suggested. Moreover, in combination with recent demonstrations that careful device construction can dramatically improve BPVE performance~\cite{Alexe11p256,Bhatnagar13p2835,Zenkevich14p161409}, these results indicate that BPVE can be much stronger than previously thought, bolstering hopes that the phenomenon can be successfully exploited.

S. M. Y. was supported by the Department of Energy Office of
Basic Energy Sciences, under grant number DE-FG02-07ER46431, and a
National Research Council Research Associateship Award at the US Naval
Research Laboratory.
F.Z. was supported by the Office of Naval Research under Grant No. N00014-11-1-0664.
A.M.R was supported by the Office of Naval Research under grant N00014-12-1-1033. Computational support was provided by the HPCMO of the DoD and NERSC of the DOE.
%merlin.mbs apsrev4-1.bst 2010-07-25 4.21a (PWD, AO, DPC) hacked
%Control: key (0)
%Control: author (8) initials jnrlst
%Control: editor formatted (1) identically to author
%Control: production of article title (-1) disabled
%Control: page (0) single
%Control: year (1) truncated
%Control: production of eprint (0) enabled
%

%\bibliography{}
\end{document}